# Prototype of the Readout Electronics for WCDA in LHAASO

Lei Zhao, *Member, IEEE*, Cong Ma, Shaoping Chu, Xingshun Gao, Zouyi Jiang, Ruoshi Dong, Shubin Liu, *Member, IEEE*, and Qi An, *Member, IEEE*

*Abstract—* In the Large High Altitude Air Shower Observatory (LHAASO), the Water Cherenkov Detector Array (WCDA) is one of the key parts. The WCDA consists of 3600 Photomultiplier Tubes (PMTs) scattered in a 90000 m$^2$ area, and both high precision time and charge measurements are required over a large dynamic range from 1 to 4000 Photo Electrons (P.E.). To achieve time measurement precision better than 500 ps RMS, high quality clock distribution and automatic phase compensation are needed among the 400 Front End Electronics (FEE) modules. To simplify the readout electronics architecture, clock, data, and commands are transferred simultaneously over 400-meter fibers, while high speed data transfer interface is implemented based on TCP/IP protocol. Design and testing of the readout electronics prototype for WCDA is presented in this paper. Test results indicate that a charge resolution better than 10% RMS @ 1 P.E. and 1% RMS @ 4000 P.E., and a time resolution better than 300 ps RMS are successfully achieved over the whole dynamic range, beyond the application requirement.

## I. Introduction

THE Large High Altitude Air Shower Observatory (LHAASO) is a large air shower particle detector array in the R&D phase, aiming at a very high energy gamma ray source survey [1]. It is proposed to be built at an altitude of more than 4000 meters above sea level. In LHAASO, the Water Cherenkov Detector Array (WCDA) [2] is one of the key detectors. It consists of four 150 m ×150 m water ponds, and 3600 Photomultiplier Tubes (PMTs) are placed under water in an area of 90000 m$^2$.

There exist several challenges in the design of the readout electronics for WCDA in LHAASO. First, both high precision time and charge measurements are required over a large dynamic range from 1 to 4000 Photo Electrons (P.E.). The requirements on the electronics are: the charge resolution is better than 30% RMS @1 P.E. and 3% RMS @ 4000 P.E., and the time resolution is better than 500 ps RMS over the whole dynamic range [3]. Second, considering the large scale of the WCDA, the Front End Electronics (FEE) modules are placed above water close to the PMTs in order to guarantee good measurement precision. Each FEE is responsible for the readout of 9 PMTs nearby, so a total of 400 FEEs are needed. Since the PMT signals are processed and digitized on front end nodes, long distance transmission of data, clock, and commands is inevitable. To achieve a good time resolution, a high quality clock has to be distributed over this large area, and automatic clock phase alignment among different FEE nodes are expected under varying ambient temperature conditions. In addition, due to the "triggerless" architecture, high speed data transfer is required based on TCP/IP protocol.

Efforts were devoted to handle the above challenges, and a prototype of the electronics was designed and tested, as presented in this paper. Section II outlines the architecture of the readout electronics for WCDA in LHAASO. In Section III, we present the analog front end circuit design, as well as the clock and data transfer interface. We conducted tests to evaluate the performance of this prototype electronics, as described in Section IV. In Section V, we conclude this paper with a summary of what has been achieved.

## II. System Architecture

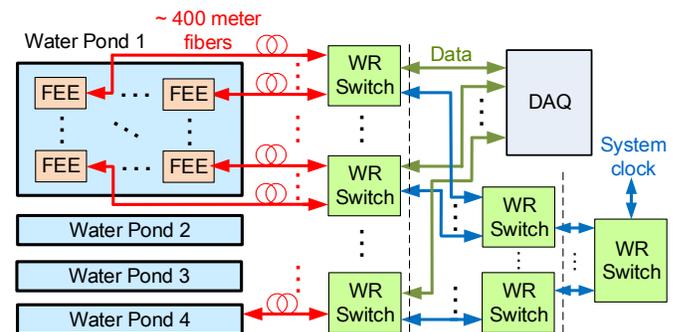

Fig. 1. Architecture of the readout electronics for WCDA.

As shown in Fig. 1, there are 100 FEEs placed above each

Manuscript received Jun. 24, 2016; revised Mar. 17, 2017.
This work was supported in part by the National Natural Science Foundation of China under Grant 11175174, in part by the Knowledge Innovation Program of the Chinese Academy of Sciences under Grant KJCX2-YW-N27, and in part by the CAS Center for Excellence in Particle Physics (CCEPP).
The authors are with the State Key Laboratory of Particle Detection and Electronics, University of Science and Technology of China, Hefei, 230026; and Modern Physics Department, University of Science and Technology of China, Hefei, 230026, China (telephone: 086-0551-63601925, corresponding author: Qi An, e-mail: anqi@ustc.edu.cn).

water pond of the WCDA. PMT output signals are digitized and processed to calculate the time and charge information. These data are transferred to White Rabbit (WR) switches and then further to the DAQ, while the system clock is also transmitted to these WR switches and further distributed to FEEs over a distance of around 400 meters. Optical fiber is employed to mitigate Electromagnetic Interference (EMI) and isolate the ground connection over long distance, to guarantee high fidelity clock distribution as in [4], [5]. Different from the systems in [4], [5], to simplify the system structure, clock, data, and commands are mixed together and transmitted through the same fiber path for each FEE. Since there is no air conditioning in the experimental hall, the ambient temperature would change from morning to night, as well as in different seasons. To address this issue, WR technique [6]–[8] is employed in the readout electronics, and WR switches are used as the bridge between FEE and DAQ, meanwhile distributing the system clock to 400 FEE nodes. The WR technique was proposed by European Organization for Nuclear Research (CERN) and Gesellschaft für Schwerionenforschung (GSI) in 2008, and it is aimed at achieving an Ethernet-based network for general purpose data transfer and clock synchronization. Based on the Precise Timing Protocol (PTP) [9], the clock phases of multiple nodes can be aligned/compensated, and a sub-nanosecond clock compensation precision (i.e. clock phase error after alignment) can be achieved [7]. Considering that the compensation precision in the electronics of WCDA in LHAASO is beyond the capability of the current WR technique, we proposed on an enhanced method based on the original WR technique [26].

Considering the complicated phenomena of cosmic ray showers, "triggerless" structure [10], [11] is adopted for a high flexibility. It means that all raw data need to be transferred to DAQ, and the trigger processing is done with software in DAQ. High speed data transfer interface based on standard TCP/IP is required to be integrated in the FEE. Detailed information about the FEE is presented in Section III.

## III. CIRCUIT DESIGN

As shown in Fig. 1, the readout electronics consists of two parts -- WR switches used as the data & clock transfer bridges and the 400 FEEs, the design of which are the kernel task in the readout electronics design.

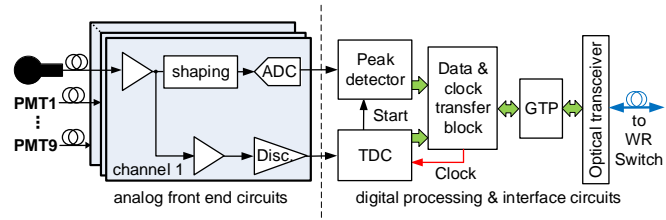

Fig. 2. Block diagram of the FEE.

Fig. 2 shows the block diagram of the FEE. In each FEE, there are two functional parts: the analog front end circuits and digital processing & interface circuits, as shown in Fig.2. The charge measurement is based on amplification, shaping, Analog-to-Digital Conversion (A/D Conversion), and peak detection, while the time measurement is achieved based on leading edge discrimination and Time-to-Digital Conversion. The Time-to-Digital Converter (TDC) and all the digital signal processing, as well as the clock and data transfer interface is integrated within one FPGA (Field Programmable Gate Array) device. The data results are sent out from GTP interface (high speed transceiver IP Core from Xilinx Inc.) in the FPGA and then converted to optical signals which are transferred to WR switches through fibers.

### A. Analog Front End Circuits

Fig. 3 shows the block diagram of the analog front end circuits for one PMT. To cover 1~4000 P.E. dynamic range, both the anode and one dynode are read out. As for charge measurement, the anode and the dynode channel share a similar structure and cover a range of 1 ~ 133 P.E. and 30 ~ 4000 P.E., respectively. The input signal from the anode is amplified by $A_1$ (AD8000 from Analog Device Inc.) and split into two paths. One is processed by a $RC^2$ shaping circuit ($R_1$, $C_1$ & $R_2$, $C_2$ and $A_2$ in Fig. 3), and then digitized by a 12 bit Analog-to-Digital Converter (ADC, AD9222 from Analog Device Inc.) with a sampling rate of 62.5 MHz. To fully utilize the Full Scale Range (FSR) of the ADC (2 $V_{pp}$ in this design), the single-ended signal is converted to a differential pair by $A_3$ in Fig. 3, with a common-mode baseline of 1 V. As for the dynode channel, since the input signal polarity is opposite to the anode, the only difference is that $A_6$ works in the non-inverting mode while $A_2$ works in the inverting mode.

The other output of $A_1$ is further amplified by $A_4$ and fed to two discriminators with two different thresholds. The high threshold is employed to avoid deterioration of time measurement resolution which would be caused by interference in the baseline of the large input signal.

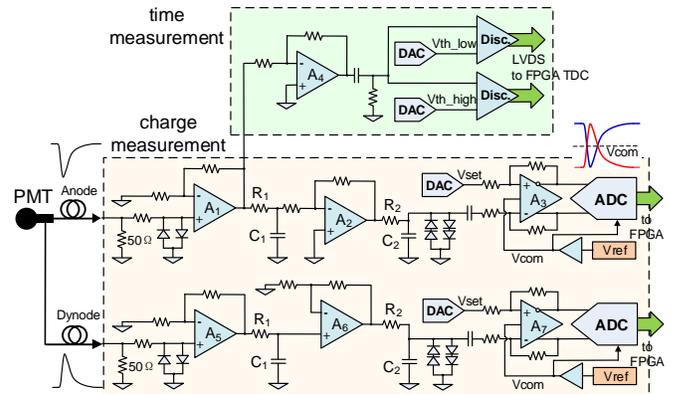

Fig. 3. Analog front end circuits for time and charge measurement.

Studies were made to optimize the circuit design parameters. In our previous research, the $RC^2$ shaping circuit structure is selected. In recent work, we made further studies on the circuit, and conducted tests to compare the actual performance with the simulation results.

Fig. 4 shows the simulation and test results of charge measurement with different time constant values ($\tau = R_1 \cdot C_1 = R_2 \cdot C_2$). The factors which contribute to the charge measurement resolution include: noise of the analog circuits before A/D conversion, error of peak detection, and the ADC noise (quantization noise etc.). As for small input signals, the noise of the analog circuits and ADC dominates the performance, while the peak detection error is the main factor with large input signals. Fig. 4 (a) and (b) show the simulation and test results with 1 P.E. and 4000 P.E. input signal amplitudes, respectively, and it indicates the test results agree with the simulations well. According to Fig. 4, a time constant of 40 ns is finally selected to make optimization for the charge measurement resolution with both small and large inputs.

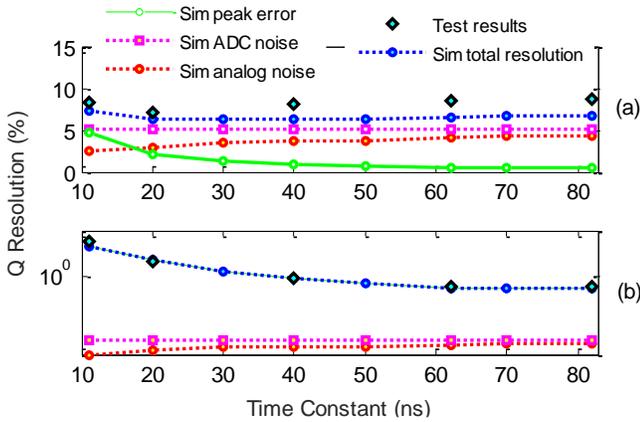

Fig. 4. Simulation and test results with different time constant values. (a) with 1 P.E. input; (b) with 4000 P.E input signal.

### B. Peak Detection Logic and TDC Design in FPGA

The outputs of the discriminators and ADCs are fed to an FPGA device (XC7A200T-FFG1156 in Artix-7 Series of Xilinx Inc.) for time and charge calculation.

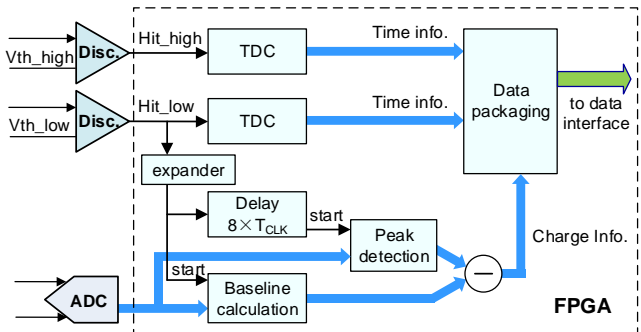

Fig. 5. Logic block diagram of the time and charge calculation.

As shown in Fig. 5, the output signal from the discriminator is used as the "start" flag for the digital peak detection logic. Since the delay of A/D Conversion is around 8 sampling clock periods [12], we calculate the averaged value of the ADC output signal baseline within this time interval. To suppress the signal baseline fluctuation, we subtract this baseline value from output of the peak detection logic to get the final charge information.

As for time digitization, two TDCs are used to process the discriminator outputs with the low and high thresholds, and thus a total of 18 TDC channels are required. Considering the requirement of time measurement (500 ps RMS), a high resolution TDC is designed based on the FPGA device. Meanwhile, the time measurement value is required to contain the information of year, month, date, hour, and second when the event occurs, and this information must be synchronized with the global UTC (Coordinated Universal Time) through the GPS (Global Positioning System) in the clock system of the whole LHAASO project. This means that this TDC should also have a very large measurement range.

As shown in Fig. 6, this TDC consists of three stages to achieve both high precision and a large measurement range. This first stage is based on multiphase clock interpolation method [13], [14]. Compared with the FPGA TDC [15]–[19] based on carry line interpolation, it features a much simpler structure, and its resolution is good enough for the application. The PLL in Fig. 6 receives the 62.5 MHz clock signal and generates four synchronized 375 MHz clock signals with 0°, 45°, 90°, and 135° phases. Through the inverter of clock buffer in each logic element, a total of 8 clock signals with a phase interval of 45° between each other can be obtained. With these multiphase clocks, a bin size of 333 ps can be achieved. To reduce consumption of the global clock resources (i.e. "BUFG" in Artix 7 FPGA [20]) and enhance the reliable working frequency of the logic, we use ISERDESE2 (close to the I/O regions in Xilinx FPGA devices) to implement these D-type Flip Flop (DFF) arrays in Fig. 6.

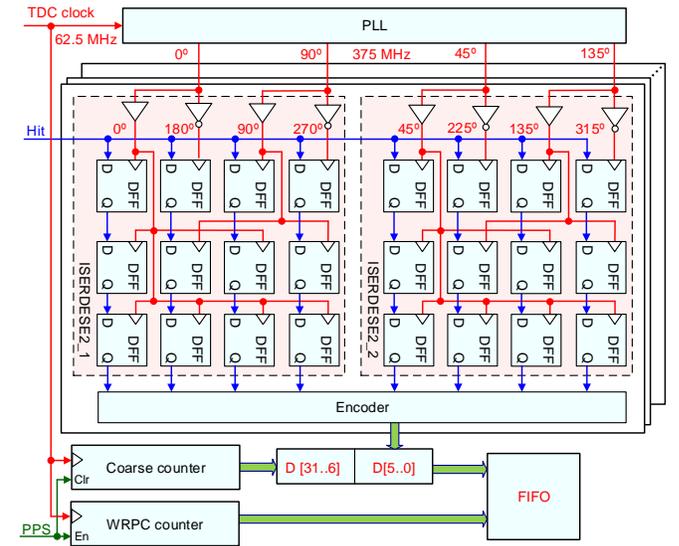

Fig. 6. Structure of the FPGA based TDC with three measurement stages.

The second stage is a coarse time counter working at the 62.5 MHz frequency. Its bin size is 16 ns, and the measurement range is up to 1 second. A larger range up to second, date, month and year is achieved using the counter in the White Rabbit PTP Core [21] (WRPC, it will be presented in the following subsection) with a bin size of 1 second. With this design, a bin size of 333 ps and a measurement range up to years is achieved.

*C. Data and Clock Transfer Interface*

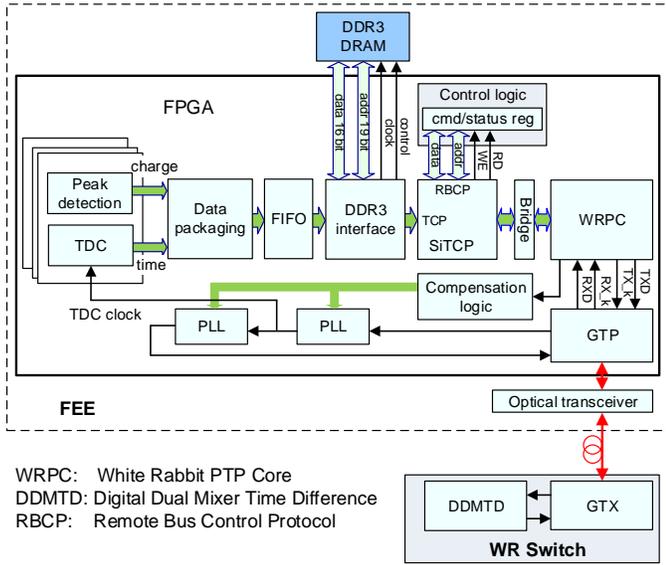

Fig. 7. Data and clock transfer interface in FPGA.

After the charge and time information is obtained, another important task is to package these data based on TCP/IP standard and send them to the DAQ through WR switches. At same time, the data & clock transfer interface of each FEE is also responsible for receiving the clock and commands from the WR switch, and aligning the clock phase with automatic, high precision phase compensation. This interface design is shown in Fig. 7.

To reduce the logic complexity and guarantee a good stability, the SiTCP core [22] is used for data packaging. To be compatible with the WR switches, the WRPC is employed to interface with the WR switch. A bridge is designed to transfer data between the GMII interface of SiTCP and the Wishbone interface of WRPC. The data are further transferred from WRPC to GTP interface of the FPGA, then pass the optical transceiver and are finally sent to the WR switch through fibers. The commands from the DAQ are sent backwards via the same path. We use the RBCP (Remote Bus Control Protocol) port of SiTCP core to send these commands to the control logic and read data from the status registers to the DAQ.

The clock is also transmitted through the optical fiber from the WR switch to FEE, and sent back to the WR switch. Attention has to be paid to the problem that delay of the clock transmission path (including the fiber as well as electronics) would change with temperature. To address this issue, the Digital Dual Mixer Time Difference measurement module (DDMTD) [23] in the switch measures the roundtrip delay variation between its output and input clock signals. FEE can compensate the clock phases according to this information. In WCDA, the clock phase compensation precision is required to be better than ±100 ps. To meet this requirement, we revised the original WR method, and enhanced its performance by designing algorithms (in "Compensation logic" in Fig. 7) for automatic delay increment assignment between the upwards and downwards paths, and moreover we used the internal PLL in the FPGA to adjust the clock phase with a step size of around 15 ps, instead of the complex external delay adjustment circuits in the traditional WR technique [24], [25]. The details are included in Ref [26].

## IV. TEST RESULTS

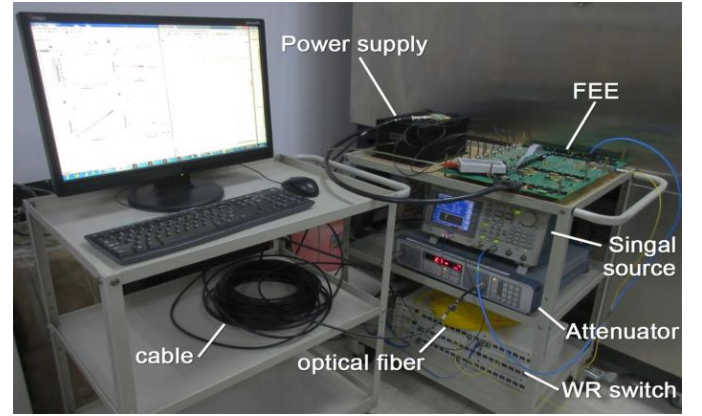

Fig. 8. System under test.

After design and fabrication of the prototype electronics, we conducted a series of tests to evaluate the performance. As shown in Fig. 8, we used an arbitrary signal source (AFG3252 from Tektronix Inc.) to generate the input signal for the FEE, according to the PMT (CR365 from Hamamatsu Inc.) output waveform recorded by a high speed oscilloscope. The signal amplitude was tuned by an external step attenuator (RSP 831.3515.02 from ROHDE & SCHWARZ Inc.), and then the time and charge measurement resolution could be tested with different input amplitudes. The output data from the FEE were transferred to WR switches and then sent to a remote PC through Ethernet. The clock signal was also transmitted to the FEE through WR switches and optical fibers.

*A. Charge and Time Measurement Performance*

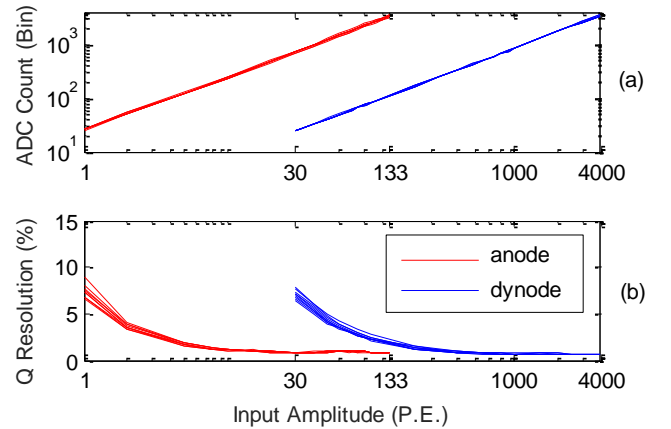

Fig. 9. Charge measurement performance test results. (a) charge measurement value versus input amplitude; (b) charge resolution versus input amplitude.

Fig. 9 shows the charge measurement performance test results of all the 9 channels in the FEE. Fig. 9 (a) is the charge measurement values with different input amplitudes, and (b) is the charge resolution. It indicates that the charge resolution is better than 10% RMS @ 1 P.E. and 1% RMS @ 4000 P.E., beyond the application requirement.

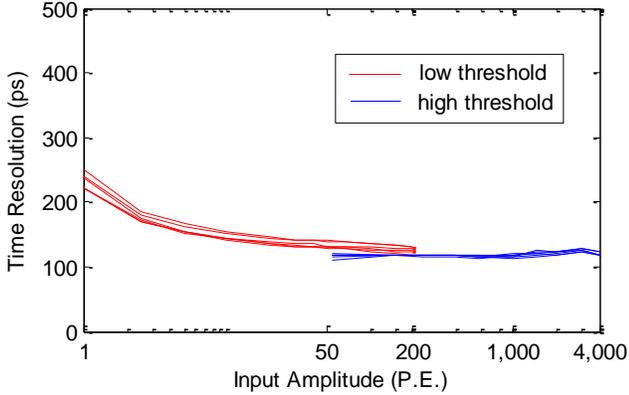

Fig. 10. Time measurement performance test results.

The time resolution test results of all the channels in the FEE are shown in Fig. 10, which indicates a time resolution better than 300 ps RMS is achieved over 1 ~ 4000 P.E. dynamic range.

### B. Clock Synchronization & Data Transfer Performance

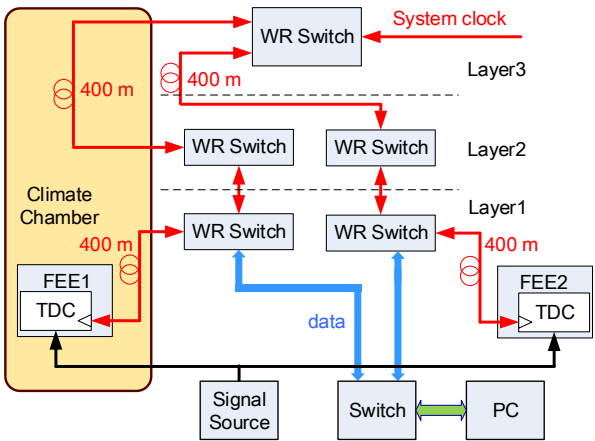

Fig. 11. Clock synchronization test with three layers of WR switches.

In WCDA readout electronics, the clock signal is required to be distributed to a total of 400 FEEs, and thus three layers of WR switches are needed. Considering the data rate limitation of WR switches, data of the FEEs are accumulated to the first layer of WR switches and sent to DAQ directly, as shown in Fig. 1 & Fig. 11. Meanwhile, the third (top) layer of WR switch receives the reference clock signal form the global clock system of LHAASO, and distributes it through the other two layers of WR switches to FEEs. To evaluate the clock compensation performance, we established a test bench, as shown in Fig. 11. In actual application, the FEEs and fibers are placed in experimental hall without air conditioning, while the WR switches are placed in the control room with a stable ambient temperature. To evaluate the effect of our compensation method in a much more severe condition, we placed one branch of the clock transmission path (including FEE1 and two 400 meter fibers) in a climate chamber, and the other branch in a constant temperature of 27 ℃.

We changed the temperature of the climate chamber from -10 ℃ to 60 ℃, and observed the variation of the time measurement difference between FEE1 and FEE2, and the clock phase compensation precision can be obtained.

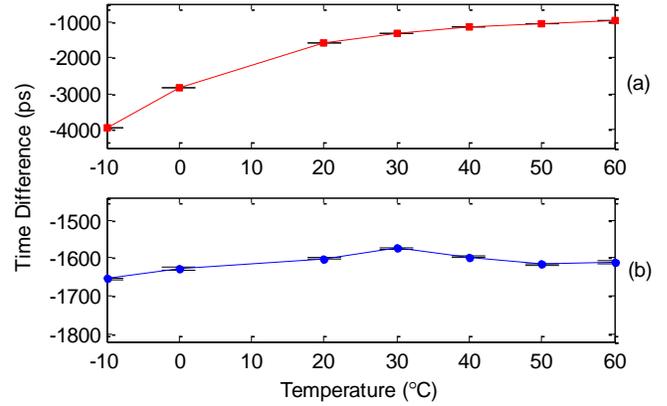

Fig. 12. Clock compensation test results. (a) without compensation; (b) with compensation.

The results in Fig. 12 indicate that the clock phase variation with temperature is reduced from 2.95 ns (Fig. 12 (a)) to within 100 ps (i.e. compensation precision), which is good enough for the application.

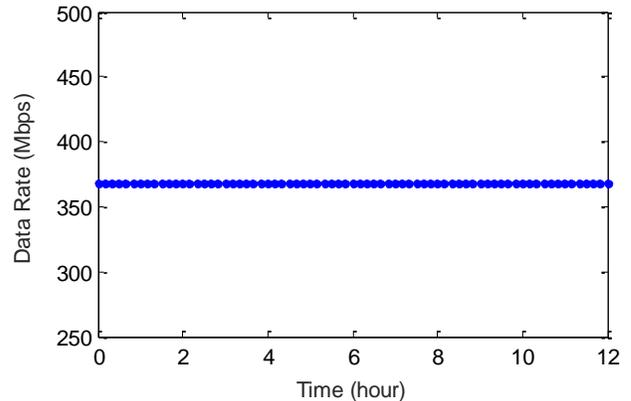

Fig. 13. Data rate test results.

We conducted initial tests on the data transfer rate, under the condition that data, clock, and commands are transferred together based on the same fiber. As shown in Fig. 13, the data rate test results over 12 hours are around 368 Mbps, which also meet the application requirement.

## C. System Temperature Drift Test

We conducted temperature drift tests on the prototype electronics.

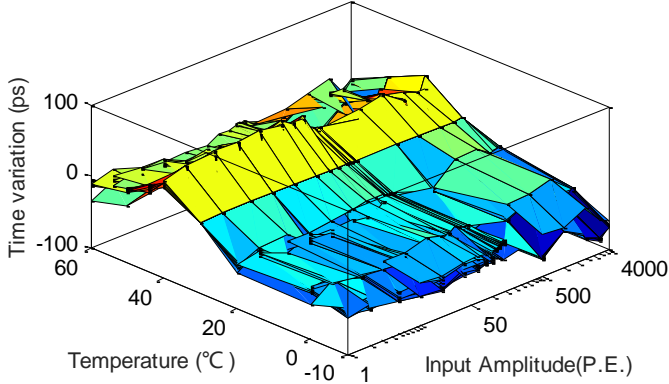

Fig. 14. Temperature drift test results of time measurement.

Fig. 14 shows the temperature drift test results of the time measurement of all 9 PMT readout channels within one FEE (the result of each channel corresponds to one layer in Fig. 14). In the input dynamic range from 1 P.E. to 4000 P.E., the time measurement result variation is within ±100 ps with temperature changing from -10 ℃ to 60 ℃, which is good enough for the application.

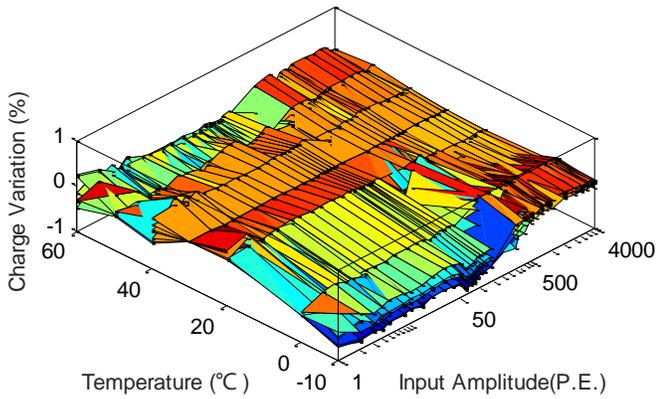

Fig. 15. Temperature drift test results of charge measurement.

Results of charge measurement temperature drift test are shown in Fig. 15 (the result of each channel corresponds to one layer in Fig. 15). The charge measurement result varies within ±1% over the whole input dynamic range, which also satisfies the application requirement.

We conducted tests on time and charge measurement resolution in different temperatures, and almost no variation was observed.

## D. Initial Integration Tests with the PMT

After electronics tests using the signal source, integration tests were conducted, by connecting the FEE with PMTs.

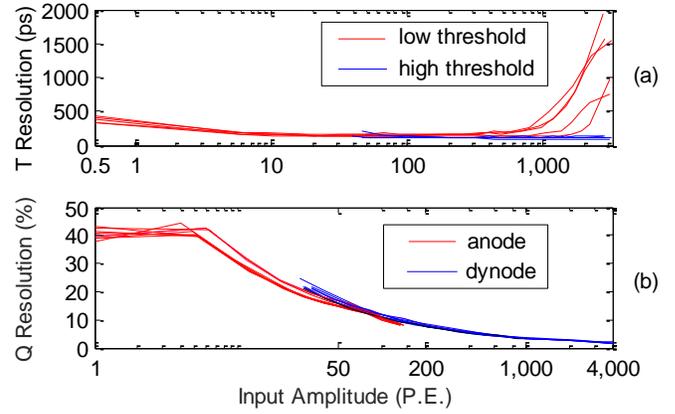

Fig. 16. Integration test results of the charge and time measurement performance. (a) time measurement performance; (b) charge measurement performance.

Fig. 16 (a) shows time measurement performance of the FEE when connected to PMTs, and a time resolution better than 500 ps RMS is achieved. Fig. 16 (a) also indicates the time resolution with the low threshold discrimination deteriorates in the range above 500 P.E., which can be improved obviously by the high threshold discrimination.

Fig. 16 (b) is the overall charge resolution of the PMT and FEE. The charge resolution of the system (including the both the FEE and PMT) is better than 50% RMS @1 P.E. and 3% RMS @4000 P.E., which meets the application requirement (50% RMS @ 1 P.E. and 5% RMS @4000 P.E.). In Fig. 16 (b), it does not exhibit a monotonic drop in the time resolution curve with input from 1 P.E. to 5 P.E. The reason is that there exists uncertainty of the photon amounts fed to PMTs from a light source, when only a few photons are generated. As for the input of 1 P.E., this uncertainty can be effectively suppressed, as the light source can be adjusted to the minimum to generate 1 photon or no photon. Even if in this situation, there still exists the possibility of 2 or 3 photons input to PMTs, as shown in the energy spectrum of Fig. 17 (Channel No. 1 in the FEE with 1 P.E. input signal). Therefore, the actual system resolution @ ~ 10 P.E. is a little better than the result in Fig. 16 (b).

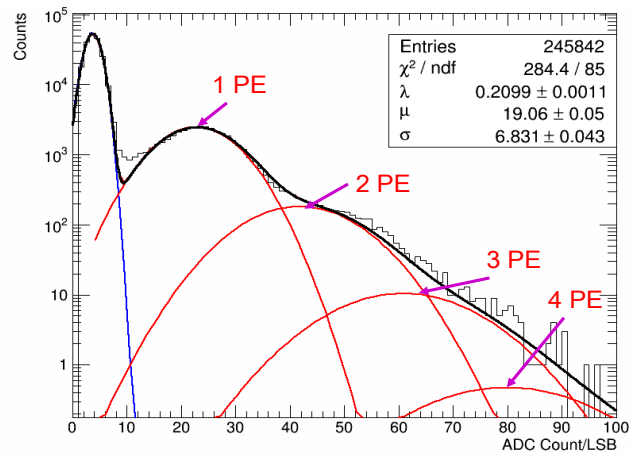

Fig. 17. Energy spectrum with the light source set to output 1 photon.

As shown in Fig. 17, by fitting the distribution of charge measurement @ 1 P.E., the charge resolution can be calculated to be around 35.8% RMS (~ $\sigma/\mu$ in Fig. 17), which is a little better than the results in Fig. 16 (b).

## V. CONCLUSION

A prototype of readout electronics was designed. Through signal processing using two channels (anode and dynode) and parameter optimization of the analog circuits, high precision measurement is achieved over a large dynamic range of 1~4000. FPGA-based TDCs using three stage time interpolation were designed to cover a large time measurement range up to years with a good resolution. Meanwhile, we proposed an enhanced method based on the WR technique to achieve a clock compensation precision better than 100 ps over a temperature range from -10 ℃ to 60 ℃. In addition, we also package data results based on the TCP/IP protocol with a high data rate, and mix data, clock, and commands together, which are transmitted within the same fiber. Tests were conducted to evaluate its performance, and the results indicate that this prototype has achieved a time resolution better than 300 ps RMS, and a charge resolution better than 10% RMS @1 P.E. and 1% RMS @ 4000 P.E., beyond the application requirement.

## ACKNOWLEDGMENT

The authors would like to thank Dr. Zhen Cao, Huihai He, Zhiguo Yao, Mingjun Chen, Cheng Li, and Zebo Tang for their kindly help. And last, the authors thank all of the LHAASO collaborators who helped to make this work possible.